\documentclass[a4paper]{article}
\usepackage[a4paper,left=1.5cm,right=1.5cm,top=2.5cm,bottom=2.5cm]{geometry}
\usepackage{hyperref}
\usepackage{multicol}
\usepackage{graphicx}
\usepackage{float}
\usepackage[utf8]{inputenc}
\usepackage[numbers]{natbib}

\title{Protein motion in the nucleus: from anomalous diffusion to weak interactions}
\date{}

\begin{document}

\author{Maxime Woringer$^{1,2,*}$, Xavier Darzacq$^{1,}$\thanks{
    Correspondence to \href{mailto:maxime.woringer@berkeley.edu}{maxime.woringer@berkeley.edu} and \href{mailto:darzacq@berkeley.edu}{darzacq@berkeley.edu}.}}

\footnotetext[1]{Department of Molecular and Cell Biology,
  Li Ka Shing Center for Biomedical and Health Sciences,
  CIRM Center of Excellence,
  University of California, Berkeley, CA 94720, USA.}
\footnotetext[2]{Unité Imagerie et Modélisation, Institut Pasteur,
    25 rue du Docteur Roux, 75015 Paris, France,
    Sorbonne Universités, CNRS, F-75005 Paris, France}

\maketitle

%% =============================
%% ========= ABSTRACT ==========
%% =============================
\begin{abstract}
  \textit{Understanding how transcription factors (TFs) regulate mammalian gene expression in space and time is a central topic in biology. To activate a gene, a TF has first to diffuse in the available space of the nucleus until it reaches a target DNA sequence or protein (target site). This eventually results in the recruitment of the whole transcriptional machinery.}

  \textit{All these processes take place in the mammalian nucleoplasm, a highly organized and dynamic environment, in which some complexes transiently assemble and break apart, whereas others appear more stable. This diversity of dynamic behaviors arises from the number of biomolecules that make up the nucleoplasm and their pairwise interactions. Indeed, interactions energies that span several orders of magnitude, from covalent bounds to transient and dynamic interactions can shape nuclear landscapes. Thus, the nuclear environment determines how frequently and how fast a TF contacts its target site, and indirectly gene expression. How exactly transient interactions are involved in the regulation of TF diffusion is unclear, but are reflected by live cell imaging techniques such as fluorescence correlation spectroscopy, fluorescence recovery after photobleaching or single-particle tracking. Overall, the macroscopic result of these microscopic interactions is almost always anomalous diffusion, a phenomenon widely studied and modeled.}

  \textit{Here, we review the connections between the anomalous diffusion of a TF and the microscopic organization of the nucleus, including recently described topologically associated domains and dynamic phase-separated compartments. We propose that anomalous diffusion found in single particle tracking (SPT) data result from weak and transient interactions with dynamic nuclear substructures, and that SPT data analysis would benefit form a better description of such structures.}
\end{abstract}
\begin{multicols}{2}

%% =================================
%% ========= INTRODUCTION ==========
%% =================================
\section{Introduction}
\begin{figure*}[!h]
\centering
\includegraphics[width=10cm]{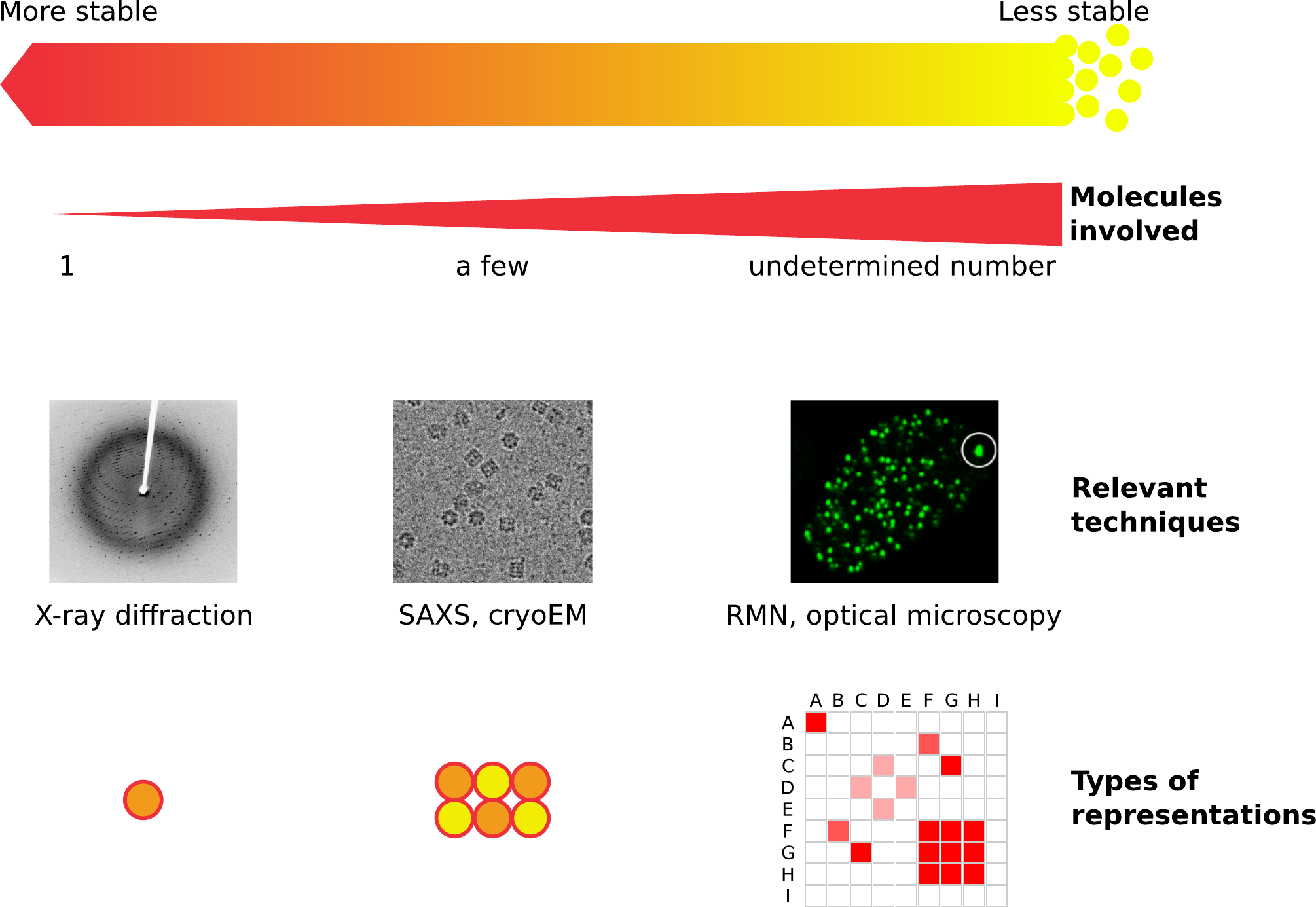}
\caption{Biological interactions cover a wide spectrum in terms of number of molecules involved and stabilities. On the one end of the spectrum, stable protein complexes very stably associate and can be purified and further imaged by techniques such as X-ray diffraction. On the other end of the spectrum, very labile, transient interactions can involve thousands of proteins in vivo, whereas none of the interactions can be captured by traditional biochemistry. (bottom row) As the valency of interactions increases from a few strongly interacting partners to many weakly interacting partners, new graphical representations are needed, since traditional schematics representing macromolecular complexes (left) cannot account for the complexity resulting from one protein weakly interacting with dozen of proteins (right). In that case, matrices of pairwise interactions between proteins A-I might be more relevant.}
\label{fig:fig1}
\end{figure*}

Mammalian gene expression and its regulation take place in the nucleus, a highly complex and sub-compartmented organelle. Interactions strengths between nuclear constituents span several orders of magnitude, from covalent bounds to "strong" non-covalent interactions. These interactions lead to the formation of macromolecular structures, either stable (Figure~\ref{fig:fig1}-left; for instance double-stranded DNA or biochemically purifiable macromolecular complexes such as the ones involved in gene expression) or transient but specific, leading to preferential associations of classes of proteins (Figure~\ref{fig:fig1}-right).

The regulation of transcription is of utmost interest as it is central not only to developmental biology, but also to cancer biology, drug screening, etc. To express a mRNA, a macromolecular complexes constituted of several subunits and dozens of proteins, the preinitiation complex, has first to assemble at the promoter of a gene in a time and space-specific manner \cite{sainsbury_structural_2015}. This complex is able to robustly integrate transient signals such as the ones mediated by proteins binding to cis-regulatory sequences such as enhancers \cite{spitz_gene_2016}.

More mechanistically, the assembly of such a complex can be characterized by a set of chemical reactions describing the progressive recruitment of factors and subunits. A kinetic rate $k$ can be associated to each of these reactions. Furthermore, traditional biochemistry and \emph{in vitro} experiments have been the methods of choice to investigate such complex processes. Most biochemical techniques involve purification steps allowing to reveal strong, non covalent interactions such as the ones occurring in a stably-assembled complex \cite{louder_structure_2016,murakami_structure_2015} (Figure~\ref{fig:fig1}-left). Then, further quantification of stoichiometry and affinity constants became possible, progressively building a network of interacting proteins, usually represented as a graph with nodes linked with arrows. 

Within this framework, the understanding of gene expression regulation reduces to elucidating how external factors (including TFs) affect the kinetic constants $k$. Although it can be assumed that kinetic rates are characterized only by the nature and concentration of enzyme, substrate and cofactors, it was shown in 1906 by Marian Smoluchowki \cite{von_smoluchowski_zur_1906} that the kinetic rate of a well-mixed reaction can be decomposed as $k=4\pi D a$. Thus, the kinetic rate $k$ is a function of both the cross-section of interaction $a$ (reflecting the chemical properties of the partners and usually studied by biochemical approaches) and the diffusion constant $D$ of the species.

Since $D$ is determined by the local environment, this finding is striking in the context of gene expression regulation: now the kinetics of one reaction depend on the whole nuclear structure. More specifically, any factor that affects diffusion in any specific or non-specific way will ultimately influence reaction rates. Indeed, interactions resulting in facilitated diffusion on a substructure (such as a TF on DNA, \cite{hammar_lac_2012,normanno_probing_2015,hansen_ctcf_2017}) or segregation inside a membrane-less compartment in a phase-separated manner \cite{strom_phase_2017} can all be seen under the unifying framework of diffusion on a surface of reduced dimensionality. Diffusion on surfaces of reduced dimensionality yields kinetics that are qualitatively different than in free, 3D diffusion and leads to potentially dramatically increased reaction rates.

Anomalous diffusion, a phenomenon occurring when a molecule explores a volume lower than predicted by diffusion, affects all proteins inside a cell. Numerous physical models can describe anomalous diffusion \cite{metzler_anomalous_2014}, and several have been applied to the motion of nuclear proteins. However, many of them only provide a phenomenological description of diffusion, rather than mechanistic insights, and radically distinct models can often fit the available data equally well.

In light of these considerations, it is worthwhile to examine the recently published discoveries describing either stable subnuclear compartments or their more transient, weak-interaction induced counterparts to highlight their influence on the diffusion of factors through dimensionality reduction. This includes TADs, LADs, nucleoli, ncRNAs, transcription factories, phase-separated domains, etc. They constitute substructures with a high valency amenable to weak interactions that can qualitatively influence diffusion and target search.

Here, we first review anomalous diffusion models applied to a protein motion and link them with a potential physical generative model. Then, we emphasize recent advances in the characterization of regions of reduced dimensionality in mammalian nuclei, both aspecific through volume exclusion and specific through transient, weak-but-specific interactions. Finally, we propose that these weak interactions shape TF dynamics, and that single particle tracking (SPT) analysis would greatly benefit from a better understanding of the pairwise interaction map between nuclear proteins.

\section{Most anomalous diffusion models reflect underlying networks of weak interactions}

The technique of choice to investigate protein motion in the nucleus of live cells is light microscopy of fluorescently tagged proteins. Different imaging and modeling modalities have been applied, including fluorescence recovery after photobleaching (FRAP), fluorescence correlation microscopy (FCS) or single-particle tracking (SPT).

In solution, the diffusion coefficient $D$ of a protein is inversely proportional to the hydrodynamic radius of the protein ($r$) and the viscosity of the medium ($\eta$) through the Stokes-Einstein relationship $D = \frac{k_B T}{6\pi\eta r}$ where $k_BT$ reflects thermal agitation, with $k_B$ the Boltzmann constant and $T$ the absolute temperature. This description, however, is too simplistic in the complex cellular environment. Indeed, with the exception of inert tracers of small molecular weight \cite{seksek_translational_1997,weiss_anomalous_2004}, it is well acknowledged that (a) macromolecules in a cell diffuse much slower than in a medium of comparable viscosity, (b) that complexes of high molecular weight can diffuse faster than small proteins, and (c) most molecules exhibit anomalous diffusion.

Thus, the diffusion of TFs cannot be described by simple friction/viscosity relationships, and their behavior, perhaps unsurprisingly, has to be seen from the angle of transient interactions with a dense matrix of interactants. In the context of this review, we define transient (or "weak") interactions as interactions that are usually too short-lived to be captured by traditional biochemistry techniques, that typically involve one or several wash step, during which proteins interacting specifically but transiently get diluted and washed out.

\begin{figure*}[h!]
\centering
\includegraphics[width=12cm]{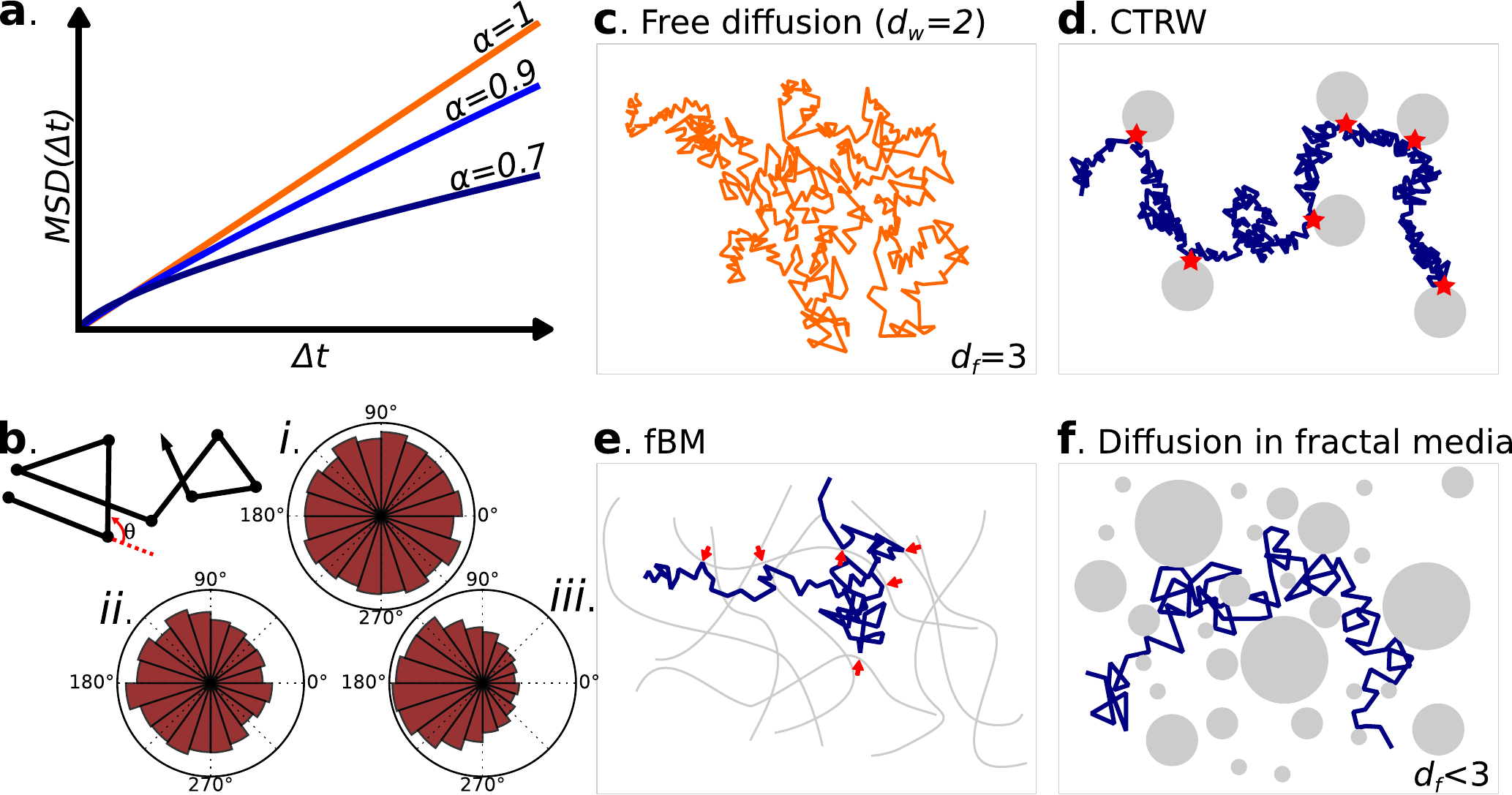}
\caption{Models of anomalous diffusion and plausible underlying physical structures. (a) and (b) characterization of anomalous diffusion. (a) A sub-linear mean-square displacement plotted as a function of time characterizes subdiffusion, and reflects how a diffusing particle explores space, the degree of anomalous diffusion is characterized by the exponent $\alpha$, the lower the $\alpha$ the more subdiffusive the process (b) An anisotropic distribution of successive angles in a walk also indicates anomalous diffusion (ii) and (iii). Although 3D free diffusion ($d_w=2$) is usually encountered in a homogenous media ($d_f=3$) (c), several types of heterogenous media can yield anomalous diffusion (d)-(f), including (d) free diffusion interspaced by long binding times –red stars–, a process called continuous time random walk –CTRW–, (e) diffusion within a viscoelastic polymer, in which a protein "bounces agains" an elastic structure and (f) diffusion within a so-called fractal media, that is a space obstructed by obstacles of all sizes.}
\label{fig:fig2}
\end{figure*}

Furthermore, diffusion of many factors is highly anomalous (more specifically, subdiffusive; Figure~\ref{fig:fig2}), meaning that the space explored over time by one factor is lower than expected by free diffusion (reviewed in \cite{hofling_anomalous_2013,metzler_anomalous_2014}). Anomalous diffusion is usually characterized by a sublinear growth of the mean squared displacement (MSD) as a function of time (Figure~\ref{fig:fig2}). Nonetheless, some anomalous diffusion processes can have a MSD identical to the MSD of free diffusion, and other characterizations are needed (Figure~\ref{fig:fig2}b). Phenomenological models have been fitted to it with success, and include continuous time random walks (CTRW; Figure~\ref{fig:fig2}d) \cite{weissman_transport_1989,saxton_biological_2007}, fractional Brownian motion (Figure~\ref{fig:fig2}e) \cite{tejedor_quantitative_2010,guigas_sampling_2008,shinkai_dynamic_2016,ghosh_fluorogenic_2017}, diffusion in fractal media (Figure~\ref{fig:fig2}f) \cite{ben-avraham_diffusion_2000,bancaud_molecular_2009,izeddin_single-molecule_2014}. Although useful as phenomenological descriptions, these models are often agnostic regarding the underlying reality of the process. In any case, the explanation of diffusion has to rely on physics and chemistry of the nucleus.

From a physical perspective, proteins can adsorb and diffuse on nuclear substructures. When this happens, the exploration properties of the protein are universally given by two parameters: first, the dimension of the random walk $d_w$, and second, the dimension of the space available to diffuse $d_f$. $d_f$  can be integer ($d_f=1$ for instance for sliding on DNA without jumps), or non-integer, a feature that characterizes self-similar structures, that is, fractals (Figure~\ref{fig:fig2}f). For the sake of this review, we will denote structures of $d_f  < 3$ as structures of \emph{reduced dimensionality}. Depending on $d_f$ and $d_w$, the motion of the protein then falls into two universal categories, termed compact and non-compact \cite{de_gennes_kinetics_1982,condamin_first-passage_2007,benichou_geometry-controlled_2010}. In a compact exploration ($d_w>d_w$), exploration is local and distance-dependent and a given site is explored repeatedly over time, in a highly recurrent manner. Conversely, in a non-compact exploration ($d_w<d_f$), the exploration is global, and every site on the structure has a constant probability to be explored (distance independence); the exploration is non-recurrent (transient). For instance, a particle freely diffusing has a $d_f$ of 2. When diffusion takes place in a 3D space ($d_f=3$) the particle tends not to never revisit sites, adopting a non-compact exploration (indeed, $d_w<d_f$). Conversely, a particle in free, Brownian diffusion ($d_w=2$) constrained to diffuse in 1D ($d_f=1$; hypothetically along a DNA fiber) will repeatedly sample the same sites (compact exploration, $d_w>d_f$). Consequently, target search times are decreased and reaction rates are increased in the compact case.

Structures of reduced dimensionality, including fractals, emerge naturally from various processes, including diffusion-limited aggregation and hierarchical assembly of macromolecular scaffolds, such as the multi-scale organization of chromatin. The goal of the next sections is to highlight a few structures of reduced dimensionality in the nucleus and how they influence kinetics of TFs.

\section{Steric hindrance in the nucleus}

Far from constituting a homogeneous medium, the nucleus is a highly organized and subcompartmentalized organelle. The main organizing structure, chromatin, constitutes approximately 10-30\% of the nuclear volume \cite{milo_cell_2016,ou_chromemt:_2017} and likely accounts for a significant part of the diffusion slowdown \cite{matsuda_macromolecular_2014}. Since every molecule has to slalom around a dense and heterogeneous chromatin environment, diffusion is impaired. Note that, however, similar diffusion coefficients are usually observed in the cytoplasm and the nucleoplasm \cite{guigas_degree_2007}, suggesting that protein crowding can also account for diffusion slowdown (\cite{ando_crowding_2010,mcguffee_diffusion_2010} and \cite{bancaud_molecular_2009} for a discussion).

\begin{figure*}[h!]
\centering
\includegraphics[width=14cm]{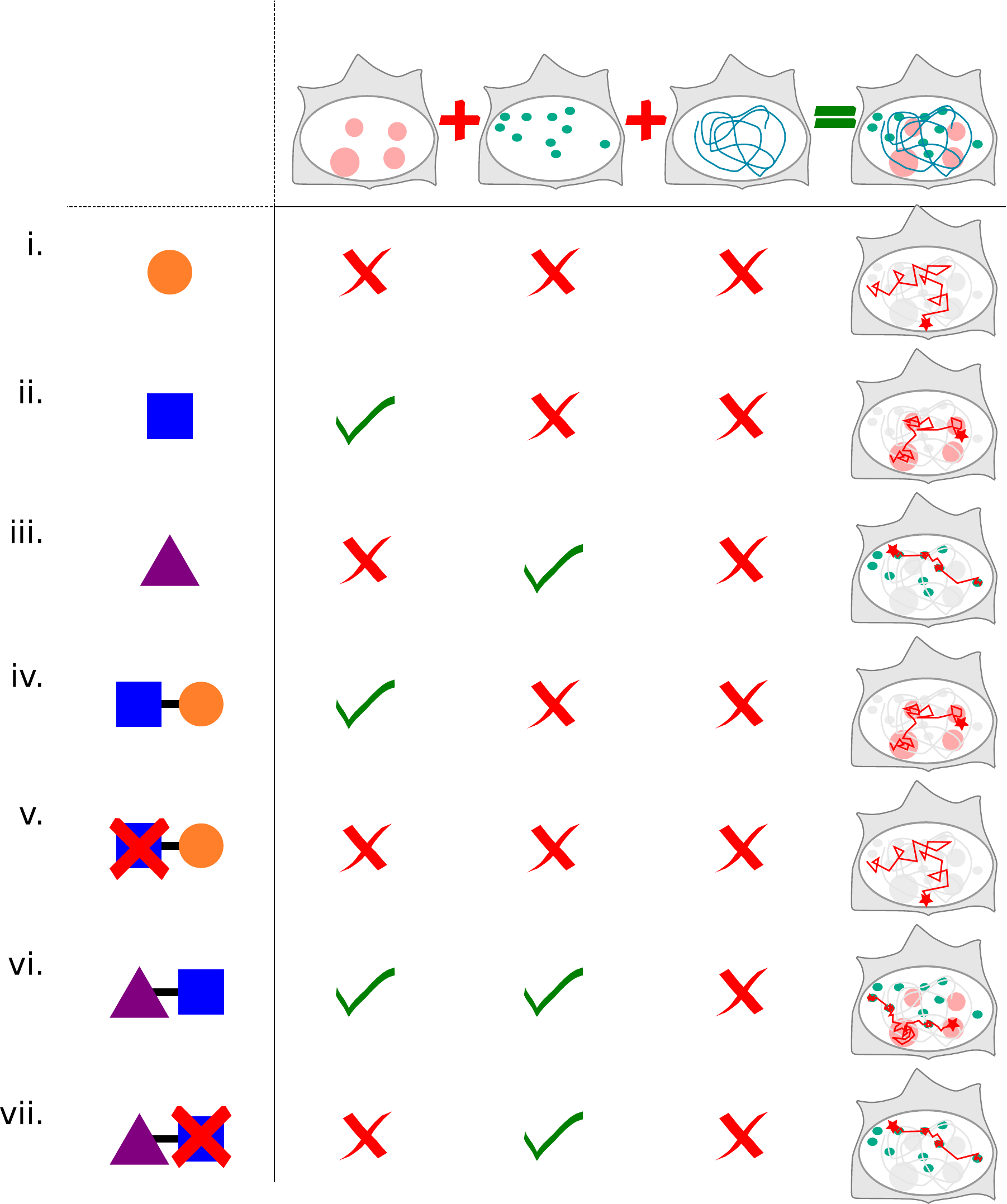}
\caption{Single particle tracking in the nucleus: weak interactions shape TF dynamics. (i)-(iii) transient interactions of individual domains. Although the round domain (i) does not interact with any particular structure (represented by the three columns of the table), (ii) the square domain interacts with a given pink structure (in column 1), and (iii) the triangle domain interacts with the small oval, green structures. All in all, this results in widely different SPT dynamics (fourth column). When domains are associated, for instance in a TF, the observed SPT is a mixture between the interactions of each single domain (iv, vi). When individual domains are mutated, the protein does not interact with a given structure anymore (v, vii), and different dynamics can be revealed.}
\label{fig:fig3}
\end{figure*}

Over the past years, organizing principles of chromatin have emerged: at large scale, the genome is segregated in chromosome territories and regions of heterochromatin/euchromatin, lamina-associated domains at the periphery and nucleoli lying more at the center. At higher magnification, chromatin is organized in areas of preferential interactions such as A/B compartments and topologically associated domains (TADs) that reflect the functional organization of chromatin (Figure~\ref{fig:fig3}a) \cite{pueschel_single_2016}.

Overall, although highly heterogeneous, chromatin in the mammalian nucleus is well described by a self-similar, fractal structure that occupies a non-zero volume. This was initially postulated \cite{grosberg_crumpled_1993}, and later evidenced by spectroscopic \cite{lebedev_fractal_2005,lebedev_structural_2008}, genomic \cite{lieberman-aiden_comprehensive_2009,grosberg_how_2012} and imaging techniques \cite{bancaud_molecular_2009,recamier_single_2014,shinkai_dynamic_2016,shinkai_bridging_2017,wang_spatial_2016}.

As a consequence, factors diffusing in the available volume are constrained by this structure \cite{goulian_tracking_2000}, possibly experiencing diffusion in a medium of reduced dimensionality, as evidenced by numerous reports \cite{bancaud_molecular_2009,hansen_robust_2018}. In a model where only volume exclusion happens, proteins of the same size and shape should have the same diffusion coefficient. Thus, the embedding structure of the nucleus only sets a lower bound on the level of anomalous diffusion that can be observed.

Several lines of argument, however, point to the fact that steric-hindrance-induced anomalous diffusion is mild. Indeed, FRAP experiments performed with protein or non-protein tracers of increasing molecular weights suggest that low molecular weight tracers diffuse almost freely in the nucleus, allowing to infer a viscosity close to the one of water \cite{seksek_translational_1997}. At higher molecular weights, anomalous diffusion becomes more and more prominent \cite{guigas_degree_2007,weiss_anomalous_2004}, eventually leading to particles being trapped in the chromatin mesh. This effect is consistent with the relatively limited volume occupied by chromatin \cite{milo_cell_2016,ou_chromemt:_2017}. Second, FRAP and FCS measurements have shown that the degree of anomalous diffusion for higher molecular weight tracers is moderate \cite{bancaud_molecular_2009}.

In conclusion, although volume exclusion by chromatin and other nuclear constituents is real, it affects all proteins of the same size in a similar manner. In contrast, a protein weakly interacting with such a structure (for instance, TFs sliding/hopping on DNA \cite{wunderlich_spatial_2008,hammar_lac_2012,coppey_kinetics_2004,iwahara_nmr_2006,doucleff_global_2008})) will immediately show a much higher level of anomalous diffusion. Furthermore, even without considering a fractal structure, simple dimensionality reduction to 1D or 2D can yield non-traditional kinetics (fractal kinetics \cite{kopelman_rate_1986,berry_monte_2002}). For instance, fractal kinetics in 2D could occur by weak interaction with the nuclear lamina, Figure~\ref{fig:fig2}b. All in all, weak and transient interactions shape the nuclear landscape and can give rise to emergent structures and properties, as exemplified in the next section.

\section{Weak interactions in the nucleus}

Unlike inert tracers whose diffusion is only determined by volume exclusion, proteins have both a relevant shape and electrostatic interaction pattern that determine their interaction landscape and thus their diffusive properties. These non-covalent interactions are obviously crucial to form biochemically stable complexes such as the transcription preinitiation complex or the spliceosome (Figure~\ref{fig:fig1}-left), but also to form dynamic emergent structures of reduced dimensionality upon which TFs can transiently adsorb and diffuse. Under this model, proteins do not form stable complexes anymore, but rather have a high number of weakly-interacting partners. The traditional representation of protein-protein interaction networks as graphs and arrows is not relevant, and can be replaced by representations such as pairwise interaction matrices (Figure~\ref{fig:fig1}-right) \cite{bergeron-sandoval_mechanisms_2016}. Indeed, simulation studies have shown that molecules can naturally undergo soft matter processes yielding structures of reduced dimensionality under very minimal hypotheses \cite{osmanovic_effect_2016,shagolsem_particle_2016}. Furthermore, the list of proteins exhibiting phase separation \emph{in vitro} or \emph{in vivo} is quickly growing, supporting the vision that the emergence of structures of reduced dimensionality is closer to a general organizing principle than an anecdotal biophysical phenomenon. Such processes include aggregation, complex coacervation, demixing and phase transition, some of them linked with transcriptional regulation \cite{tsai_nuclear_2017,kwon_phosphorylation-regulated_2013,sherry_control_2017,chong_dynamic_2017}.

First, structures of reduced dimensionality, such as aggregates, phase separated domains or subnuclear compartments require at least one \emph{multivalent partner}, that can nucleate the aggregation. As such, many abundant constituents of the mammalian nucleus have been shown to nucleate a structure of reduced dimensionality, in a manner very much akin to heterogeneous catalysis in chemistry \cite{woringer_geometry_2014}. These constituents include low complexity protein domains, that constitute the majority of the mammalian proteome \cite{shammas_mechanistic_2017,kato_cell-free_2012}, especially TFs \cite{liu_intrinsic_2006}, repeated DNA \cite{nott_phase_2015} or RNA sequences \cite{jain_rna_2017,li_phase_2012,molliex_phase_2015} or small amphiphilic molecules \cite{patel_atp_2017,patel_liquid--solid_2015}.

Second, the partners have to exhibit compatible interactions: it is chemically unlikely that both highly charged and hydrophobic proteins will coexist in the same structure without the help of additional compounds acting as counterions \cite{pak_sequence_2016}, setting the basis of a “grammar of interactions” \cite{gimona_protein_2006}, that is being progressively deciphered \cite{brady_structural_2017,reichheld_direct_2017,das_relating_2015,sherry_control_2017,patel_liquid--solid_2015,quiroz_sequence_2015}.

Third, structures of reduced dimensionality emerging from weak interactions exhibit the following properties: (1) they usually exist as an extremely dynamic equilibrium rather than a stable structure \cite{strom_phase_2017,wei_phase_2017,molliex_phase_2015}, and can thus be at the same time prevalent in the nucleus and hard to purify by traditional biochemistry that preferentially capture stable interactions. (2) Moreover, they emerge from a dynamic mesh of pairwise chemical interactions. They can show a high level of specificity, and several structures of reduced dimensionality can coexist in the same nucleus without intermixing \cite{chong_dynamic_2017,shav-tal_dynamic_2005,pak_sequence_2016,kwon_phosphorylation-regulated_2013,nott_phase_2015}. Furthermore, the number and spatial relationships of such structures is only limited by the combinatorics of chemical interactions. (3) Finally, these structures can be regulated by the well-studied post-translational machinery of eukaryotic cells. For instance, phosphorylation of one of the proteins involved in such structure can trigger the timely disassembly of the whole structure and free all the factors interacting with it \cite{cho_super-resolution_2016,cisse_real-time_2013,kwon_phosphorylation-regulated_2013}. All those factors will then exhibit a dramatically different dynamics and target search properties, potentially switching from a compact exploration mode to a non-compact one. As such, a specific (and potentially functional) group of factors can be regulated at once by modulation of the post-translational modifications of one “architectural” protein \cite{li_phase_2012,nott_phase_2015}.

The characterization of structures of reduced dimensionality emerging from weak interactions is still in its infancy, but appears more and more strongly as a clear organizing principle of mammalian nuclei. These structures create the matrix upon which fast-diffusing factors can specifically and transiently bind, diffuse and unbind, thus dynamically shaping the “diffusion landscape” of the whole transcriptional machinery.

Even though live imaging approaches specifically characterize the behavior of one single factor, they are blind to all these substructures. Indeed, SPT reflects the dynamics of proteins transiently interacting with those structures of reduced dimensionality and one TF potentially visits several of them in the span of a few tens of milliseconds. Such complex behavior therefore appears macroscopically as various kinds of anomalous diffusion.

\section{Perspectives: seeing beyond the dots}
In a complex mammalian nucleus, the diffusion of a TF is ruled by transient interactions with underlying structures of reduced dimensionality, such as detailed in the two previous sections. From a more general perspective, the question arises of how gene expression regulation processes relate to the multiplicity of structures of reduced dimensionality?

Proteins often harbor several domains, holding the potential to interact alternatively and repeatedly with multiple classes of structures of reduced dimensionality. Thus, depending on its interaction domains, a TF will “see” a different landscape and will interact with some structures whereas other factors will either be excluded or cross them without any additional interactions than limited steric hindrance (Figure~\ref{fig:fig3}). In this respect, the nucleus can be described as a “multiverse”, in which some factors coexist in the same physical space but exhibit radically distinct dynamics and interactions (Figure~\ref{fig:fig3}).

Furthermore, structures of reduced dimensionality have been proven to be functionally relevant. For example, the dynamic and regulated switching of a TF between structures of reduced dimensionality determines its function. It has been shown that TF exhibit radically different dynamics before/after a post-translational modification \cite{loffreda_live-cell_2017}, or an artificial deletion of a domain (Figure~\ref{fig:fig3}), \cite{hansen_ctcf_2017,elf_probing_2007,mazza_benchmark_2012,sekiya_nucleosome-binding_2009,claus_dna_2017,zhen_live-cell_2016}. In that case, the observed diffusion will be arising from the remaining interactions from the other interaction domains, or ultimately from simple volume exclusion \cite{isaacson_influence_2011}.

Although theoretical and experimental support for the importance of weak interactions as an architectural principle of the nucleus and gene expression regulation is being actively investigated, several questions remain unaddressed:

First, how many distinct types of structures of reduced dimensionality exist? Since the numbers of types of low-complexity domains is likely to be limited, one can expect that a limited number of such structures actually coexist at a given time in a nucleus \cite{das_relating_2015}. This implies that the SPT dynamics of TFs will fall in a limited number of categories, which in turn is determined by their combinatorial interactions with one or several of these structures. To take into consideration such processes paves the way the way for a higher-order understanding of gene expression regulation and key transitions occurring for instance during mitosis or development.

Second, can we determine the pairwise interaction matrix between low-complexity protein domains, which would allow to derive predictive dynamics of a given TF modification? Ideally, such matrix will encompass all known low-complexity domains, but also abundant multivalent RNAs and DNA sequences, and each element of this matrix will reflect the affinity between two domains under physiological conditions (Figure~\ref{fig:fig1}).

Third, how much detail is required to describe these structures of reduced dimensionality? Is the pairwise interaction between protein domains a good approximation of the properties of the nucleus? Conversely, one can imagine substructures of reduced dimensionality arising from interactions more complex than simple pairwise-interactions. Indeed, it is widely known that cooperativity plays a role in the assembly of many more or less stable macromolecular structures \cite{chronis_cooperative_2017}, including some phase-separated domains \cite{pak_sequence_2016}.

Fourth, how do key biological transitions such as differentiation intertwine with these structures of reduced dimensionality? In a similar way as pluripotency or cell-type specific TF networks have been identified, can pluripotency or cell-type specific structures of reduced dimensionality be evidenced, integrating the expression levels of TFs and providing a framework to better understand such key processes?

To answer those questions, our understanding of nuclear processes need to be drastically expanded. Hitherto, a dynamic picture of spatially segregated factors, together with their interaction matrix, is currently missing. Promising tools to access those parameters include quantitative FRET \cite{sukenik_weak_2017}, in cell NMR \cite{maldonado_-cell_2011,freedberg_live_2014,theillet_structural_2016}, low-photons SPT \cite{balzarotti_nanometer_2017}, tracking FCS \cite{limouse_intramolecular_2017}, spatially resolved FCS \cite{singh_3d_2017} and computational methods \cite{quiroz_sequence_2015,harmon_gadis:_2016}.

\section{Conclusion}
Although the so far identified key players in gene expression regulation are biochemically stable complexes that can be purified using traditional methods, increasing evidence suggest that higher-order, weaker-interaction structures, acting as structures of reduced dimensionality, play a central role in transcriptional regulation. They do so by providing a remarkably versatile way of specifically and timely regulating TF target search dynamics and thus gene expression. All in all, the functional properties of the nucleus emerge more and more as a continuum of weak, seemingly random interactions, rather than from an unstructured assembly of structured macromolecular complexes. In this context, the saying from Heraclitus makes probably more sense than ever: "The fairest order in the world is alike a heap of random sweepings".

\section{Acknowledgements}
We apologize to our colleagues whose work could not be cited due to limited space. We thank Elena Rensen and Anders S. Hansen for critical feedback on the manuscript. Work in the Darzacq lab is supported by National Institutes of Health (UO1-EB021236) and California Institute for Regenerative Medicine (LA1-08013).We are very thankful to the Sci-Hub database, without which the writing of this review would have been impossible.

%% ===============================
%% ========= REFERENCES ==========
%% ===============================
\bibliographystyle{unsrtnat}
\footnotesize\bibliography{bibliography}

\end{multicols}

\end{document}